\documentclass{PoS}
\usepackage{times} 
\usepackage{graphics}
\usepackage{epsfig}
\usepackage{amsmath,amsfonts,amssymb,amsthm}
\usepackage{booktabs}
\usepackage{slashed}

\newcommand{\bq}{ \begin {equation} }
\newcommand{\eq}{\end{equation}}
\newcommand{\ba}{\begin{eqnarray}}
\newcommand{\ea}{\end{eqnarray}}

\newcommand{\be}{\begin{equation}}
\newcommand{\ee}{\end{equation}}

\newcommand{\bea}{\begin{eqnarray}}
\newcommand{\eea}{\end{eqnarray}}

\newcommand*{\CO}{\mathcal{O}}
\newcommand*{\CI}{\mathcal{I}}	
\newcommand*{\R}{\mathbb{R}}      
\newcommand*{\CU}{\mathcal{U}}
\newcommand*{\CF}{\mathcal{F}}

\renewcommand{\figurename}{\bf Fig.}
\renewcommand{\tablename}{\bf Tab.}

\def\mathswitch#1{\relax\ifmmode#1\else$#1$\fi}
\def\mathswitchr#1{\relax\ifmmode{\mathrm{#1}}\else$\mathrm{#1}$\fi}

\newcommand{\re}{\Re e \,}

\newcommand{\dd}{{\mathrm d}}

\title{
{
\flushright{
\small \tt DESY 18-121
\\
\small  \tt KW 18-008
\\[8mm]
}
}
MBnumerics: Numerical integration of Mellin-Barnes integrals in physical regions}

\ShortTitle{Numerical evaluation of Mellin-Barnes integrals}

\author{\speaker{Johann Usovitsch}
\\
        Hamilton Mathematics Institute, Trinity College Dublin,\\
	College Green, Dublin 2, Ireland\\
        E-mail: \email{usovitsj@maths.tcd.ie}}

\author{Ievgen Dubovyk\\
       II. Institut f\"ur Theoretische Physik, Universit\"at Hamburg, 22761 Hamburg, Germany\\
       DESY Deutsches Elektronen-Synchrotron, 15738 Zeuthen, Germany\\
       E-mail: \email{e.a.dubovyk@gmail.com}}
        
\author{Tord Riemann\\
       University of Silesia, 40-007 Katowice, Poland\\
       DESY Deutsches Elektronen-Synchrotron, 15738 Zeuthen, Germany\\
       E-mail: \email{tordriemann@gmail.com}}

\abstract{We introduce techniques to treat numerically Mellin-Barnes integrals in physical regions, which arise in the need of the computation of Feynman integrals for the electroweak two-loop corrections to the pseudo observables at the Z-boson resonance.}

\FullConference{Loops and Legs in Quantum Field Theory (LL2018)\\
		29 April 2018 - 04 May 2018\\
		St. Goar, Germany}
		
\begin{document}
\section{Introduction}
Our starting point will be the loop-momenta integral representation of a scalar Feynman integral:
\begin{equation}
 G_{L}=\int\prod_{j=1}^L \frac{{\mathrm d}^{D} k_{j}}{i\pi^{D/2}}\;
 \frac{
 1
 }
 {P_{1}^{\nu_{1}}\dots P_{N}^{\nu_{N}}}.
 \label{eq:tensorIntegral}
\end{equation}
The functions $P_{i}^{\nu_i}$ in the denominator are expressed in terms of the $L$ loop-momenta $k_{l}$ which are not fixed through momentum conservation at each vertex and the $E$ linearly independent external momenta $p_{e}$:
\begin{equation}
P_{i}
=\left(\sum\limits_{l=1}^{L}a_{il}k_{l}+\sum\limits_{e=1}^{E}b_{ie}p_{e}\right)^{2}-m_{i}^{2}+i\delta,\;a_{il},\;b_{ie}\in\{-1,0,1\},
\label{eq:propagator}
\end{equation}
where the $m_{i}$ denote the masses of the corresponding virtual particles. The $i\delta$ is the Feynman prescription. In the most general case the $P_{i}$ are a linear combination of $N$ linearly independent scalar products depending on the loop-momenta $k_{l}$. The propagator exponents $\nu_{i}$ are complex variables if not stated otherwise.
Within dimensional regularization, $D=4-2\epsilon$ denotes the dimension of space-time. As
usual $D\not=4$ is used to regularize infrared or ultraviolet
divergences.

Before evaluating these integrals one often applies the Feynman trick:
\begin{equation}
 \frac{(-1)^{\nu}}{\prod\limits_{j=1}^{N}(-P_{j})^{\nu_{j}}}=\frac{(-1)^{\nu}\Gamma(\nu)
 \left(\prod\limits_{j=1}^{N}\tilde n_{j}\right)
 \delta(1-{\sum\limits_{j=1}^{N_{G}}}x_{j})}{( -k_{l}^{\mu}M_{ll'}k_{l'\mu}+2k_{l}^{\mu}Q_{l\mu}+J-i\delta)^{\nu}},\;\;\nu=\sum\limits_{j=1}^{N}\nu_{j},
 \label{eq:FeynmanIntroduction}
\end{equation}
where
\begin{equation}
M_{ll'}=\sum\limits_{j=1}^{N}a_{jl}a_{jl'}x_{j}
\label{eq:matrixM}
\end{equation}
is an $L\times L$ symmetric matrix, 
\begin{equation}
Q_{l}^{\nu}=-\sum\limits_{j=1}^{N}x_{j}a_{jl}\sum\limits_{e=1}^{E}b_{je}p_{e}^{\nu}
\label{eq:vectorQ}
\end{equation}
is a vector with $L$ components and 
\begin{equation}
J=-\sum\limits_{j=1}^{N}x_{j}(\sum\limits_{e=1}^{E}b_{je}p_{e}^{\mu}\sum\limits_{e'=1}^{E}p_{e'}^{\nu}b_{je'}g_{\mu\nu}-m_{j}^{2}),
\label{eq:scalarJ}
\end{equation}
where $x_{j}$ are the Feynman parameters introduced with the Feynman trick. The set of  Feynman parameters $\{x_{1},\dots,x_{N_{G}}\}$ corresponds to the set of functions $\{P_{1},\dots,P_{N_{G}}\}$ with positive $\{\nu_{1},\dots,\nu_{N_{G}}\}$ in Eq.\eqref{eq:tensorIntegral}. The metric tensor is $g_{\mu\nu}= \mathrm{diag}(1,-1,\dots,-1)$. The $\tilde n_{j}$ is defined as:
\begin{equation}
\tilde n_{j}\phi(\vec x)=\begin{cases}
 \int\limits_{\{x_{j}\geq0\}}\frac{\dd x_{j}\,x_{j}^{\nu_{j}-1}}{\Gamma(\nu_{j})}\phi(\vec x),\;\;\;\nu_{j}\neq -m,\\
 (-1)^{\nu_{j}}\phi^{(-\nu_{j})}(0,x_{i\neq j}),\;\;\;\nu_{j}=-m,
 \end{cases}
 \;\;\;m\in \mathbb{N}_{0},
\end{equation}
where $\phi^{(-\nu_{j})}(0,x_{i\neq j})$ means to take $(-\nu_{j})$ derivative in $x_{j}$ and then set $x_{j}$ to zero.

The Feynman integral can now be written in the Feynman parameter integral representation:
\begin{equation}
G_{L}=(-1)^{\nu}\Gamma(\nu-LD/2)
\left(\prod\limits_{j=1}^{N}\tilde n_{j}\right)
\delta(1-\sum\limits_{j=1}^{N_{G}}x_{j})\frac{\CU(x)^{\nu-(L+1)D/2}}{\CF(x)^{\nu-LD/2}},
\label{eq:finalFeynman}
\end{equation}
where
\begin{eqnarray}
 \CU(x)&=&\det M,
 \label{eq:Upolynom}\\
 \CF(x)&=&\CU(x)(Q_{l}^{\mu}M_{ll'}^{-1}Q_{l'\mu}+J-i\delta).
 \label{eq:Fpolynom}
\end{eqnarray}
From these definitions it follows that the functions $\CF(x)$ and $\CU(x)$ are homogeneous in the Feynman parameters $x_{i}$. The function $\CU(x)$ is of degree $L$ and the function $\CF(x)$ is of degree $L+1$. The functions $\CU(x)$ and $\CF(x)$ are also known as Symanzik polynomials.

\subsection{Mellin-Barnes integral}

Feynman integrals may be infrared and ultraviolet divergent. To treat these integrals in a consistent and automated way two methods are known: The Mellin-Barnes integral approach \cite{Smirnov:1999gc,Tausk:1999vh,Heinrich:2004iq,Czakon:2004wm,Czakon:2005rk,Anastasiou:2005cb,Gluza:2007rt,Gluza:2010rn,Dubovyk:2015yba,Dubovyk:2016ocz,Prausa:2017frh} and the sector decomposition approach \cite{Hepp:1966eg,Binoth:2000ps,Binoth:2003ak,Binoth:2004jv,Denner:2004iz,Heinrich:2008si}.

To derive a Mellin-Barnes integral one will use either the loop-by-loop approach \cite{Gluza:2007rt} or the global approach \cite{Dubovyk:2015yba}. Both techniques apply in their core to the $\CF(x)$ and $\CU(x)$ functions in Eq.~\eqref{eq:finalFeynman} the Mellin-Barnes integral master formula:
\begin{equation}
\frac{1}{(a+b)^\nu}=\int\limits_{-i\infty}^{i\infty} \frac{ {\mathrm d} z}{2\pi i}\,  \frac{a^{z}b^{-z-\nu}\Gamma(-z)\Gamma(\nu+z)}{\Gamma(\nu)},\quad |\arg a-\arg b| < \pi,
\end{equation}
until the integrations over the Feynman parameters can be all carried out in terms of Euler's Beta-functions:
\begin{equation}
B(\xi,\chi)=\int\limits_{0}^{\infty}\frac{x^{\xi-1}}{(1+x)^{\xi+\chi}}{\mathrm d} x=\frac{\Gamma(\xi)\Gamma(\chi)}{\Gamma(\xi+\chi)},\quad \re \xi>0,\;\re \chi>0.
\end{equation}
These steps lead to Mellin-Barnes integrands depending on a ratio of Euler's Gamma-functions $\Gamma$ depending on the integration variables $z_{i}$ and some kinematics raised to the powers of $z_{i}$.

\begin{figure}[htpb]
\centering
   \begin{center}
\includegraphics[width=0.48\linewidth]{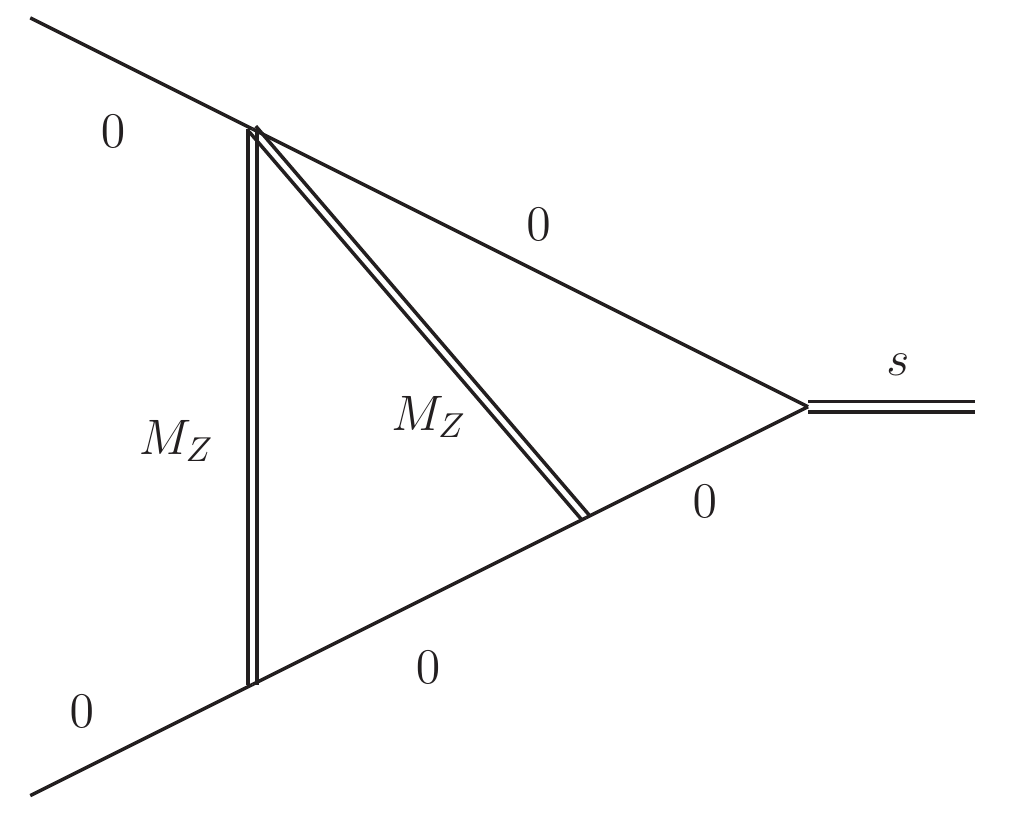}
   \end{center}
  \caption{Two-loop vertex Feynman integral with two internal massive lines and the kinematics are $p^{2}_{1,2}=0$ and $2p_{1}p_{2}=s$. The $Z$-boson mass $M_{Z}$ indicates massive propagators.}
\label{fig:0h0w14r}
\end{figure}

As an example we study the Feynman integral in Fig.~\ref{fig:0h0w14r}, whose loop-momenta integral representation contains one nontrivial numerator $k_{1}p_{1}$:
\begin{equation}
 I_{\text{0h0w14r}}= \int \frac{{\mathrm d}^{D} k_{1}}{i\pi^{D/2}} \frac{{\mathrm d}^{D} k_{2}}{i\pi^{D/2}}\,\frac{k_{1}p_{1}}{k_{1}^{2}((k_{1}-k_{2})^{2}-M_{Z}^{2})k_{2}^{2}((k_{2}+p_{1})^{2}-M_{Z}^{2})(k_{1}+p_{1}+p_{2})^{2}},
\end{equation}
and its Mellin-Barnes integral representation is:
\begin{eqnarray} 
I_{\text{0h0w14r}}=&
\int\limits_{-\frac{1}{3}-i\infty}^{-\frac{1}{3}+i\infty}\frac{{\mathrm d} z_{1}}{2\pi i}\int\limits_{-\frac{1}{3}-i\infty}^{-\frac{1}{3}+i\infty}\frac{{\mathrm d} z_{2}}{2\pi i}\,\frac{\Gamma (-z_{1}) \Gamma (-z_{2}) \Gamma (z_{2}+1)
   \Gamma (-\epsilon-z_{1}) \Gamma (2 \epsilon+z_{1}+1)
   \left(-\frac{M_{Z}^2}{s}\right)^{z_{1}}
   }
   {2 \Gamma (1-z_{2}) \Gamma
   (-3 \epsilon-z_{1}+2) \Gamma (-2 \epsilon-2
   z_{1}-z_{2})}\notag\\
   &\times\left(-s\right)^{-2 \epsilon}\Gamma (-2 \epsilon-z_{1}-z_{2})^2 \Gamma
   (-\epsilon-z_{1}-z_{2}) \Gamma
   (\epsilon+z_{1}+z_{2}+1),
   \label{eq:MBexample}
\end{eqnarray}
where $\epsilon=(4-D)/2$.
\begin{figure}[tpb]
\centering
   \begin{center}
\includegraphics[width=0.48\linewidth]{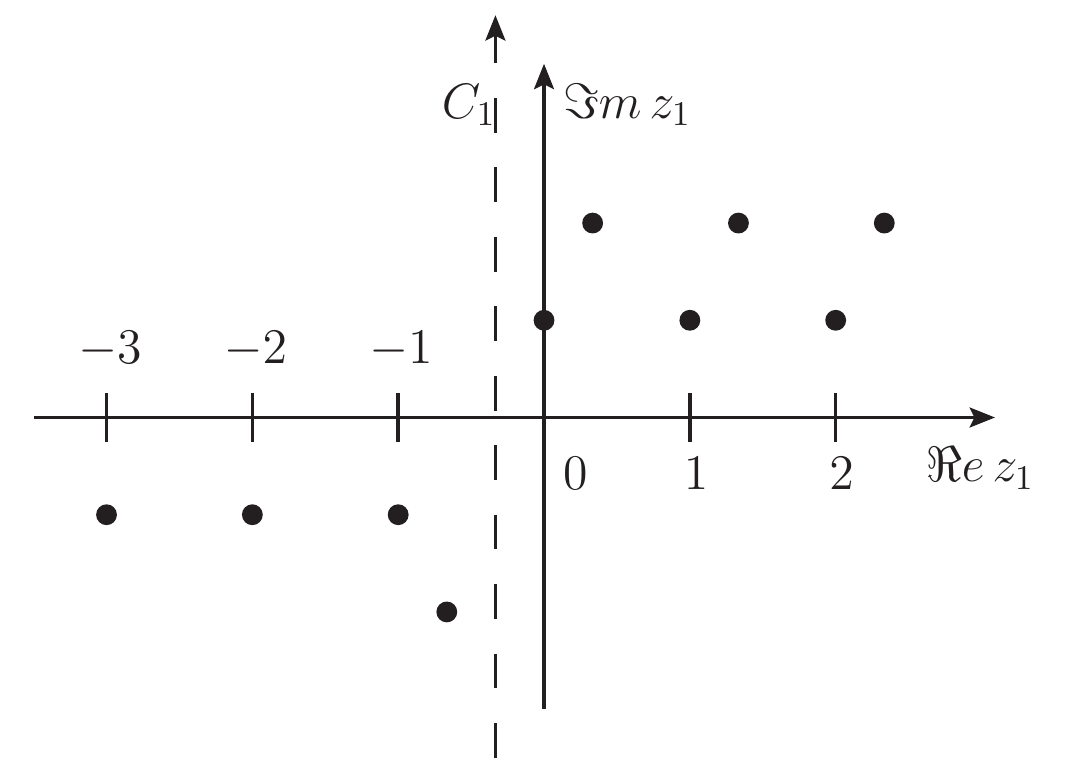}
   \end{center}
  \caption{The black dots are the poles of the integrand in Eq.~\eqref{eq:MB0h0w14finitelinearTransformation} in the $z_{1}$ complex plane. The dashed line is the integration contour parallel to the imaginary axis.}
\label{fig:contour}
\end{figure}
A straight integration contour parallel to the imaginary axis is chosen, such that all poles are well separated, see Fig.~\ref{fig:contour} for $z_{1,2}=-1/3+i t_{1,2}$, $t_{1,2}\in [-\infty,\infty]$. The expansion around $\epsilon=0$ in Eq.~\eqref{eq:MBexample} leads to a finite contribution in the lowest order:
\begin{eqnarray} 
I_{\text{0h0w14r}}=&-\int\limits_{-\frac{1}{3}-i\infty}^{-\frac{1}{3}+i\infty}\frac{{\mathrm d} z_{1}}{2\pi i}\int\limits_{-\frac{1}{3}-i\infty}^{-\frac{1}{3}+i\infty}\frac{{\mathrm d} z_{2}}{2\pi i}\,
\frac{\Gamma (-z_{1})^2 \Gamma (z_{1}+1) \Gamma (-z_{2})
   \Gamma (z_{2}+1)
   \left(-\frac{M_{Z}^2}{s}\right)^{z_{1}} \Gamma
   (-z_{1}-z_{2})^3}{2 \Gamma
   (2-z_{1}) \Gamma (1-z_{2}) \Gamma (-2
   z_{1}-z_{2})}\notag\\
   &\times\Gamma (z_{1}+z_{2}+1)+\CO(\epsilon).
   \label{eq:MBexampleseries}
\end{eqnarray}

\subsection{Minkowskian kinematics}
Whether we derive the Mellin-Barnes integrals with the loop-by-loop or the global approach, we face problems in the numerical treatment of these integrals in Minkowskian kinematics. To illustrate this we apply the well known Stirling approximation formula
\begin{equation}
\Gamma(z)\underset{|z|\to\infty}{\approx}z^{z-1/2}e^{-z}\sqrt{2\pi},\quad|\arg{z}|<\pi,
\label{eq:stirling}
\end{equation}
to the integrand in Eq.~\eqref{eq:MBexampleseries} and examine the asymptotic behavior for 
$z_{1}=-\frac{1}{3}+it_{1}$, and $z_{2}=-\frac{1}{3}+it_{2}$, $t_{1}\to-t$ and $t_{2}\to t$:
\begin{equation}
 \CI_{\text{0h0w14r}}\underset{t\to\infty}{\approx} t^{-2+2x_{1}+2x_{2}}|_{x_{1}=x_{2}=-1/3}.
\end{equation}
In comparison to the Euclidean kinematics, where the asymptotic behavior is everywhere exponentially damped, we see that for Minkowskian kinematics the asymptotic behavior is polynomial. In the case of a Mellin-Barnes integral this polynomial asymptotic behavior leads to numerous numerical instabilities, some of which are:
\begin{itemize}
 \item Oscillations are less damped compared to the Euclidean case.
 \item Integrals may be not absolutely convergent if the asymptotic behavior is worse than $1/t^{a}$, with $a<2$.
 \item At any level of accuracy, we need to evaluate the integrands for bigger values $t_{i}$ than in the case of Euclidean kinematics.
 \item In particular, if we are interested in high accuracy results, we have to evaluate the $\Gamma$ functions for very big arguments and this leads again to numerical instabilities.
\end{itemize}

\section{Techniques to treat Mellin-Barnes integrals in Minkowskian kinematics}
We assume that the treatment of one-dimensional Mellin-Barnes integrals is a solved problem by means of contour deformation \cite{Freitas:2010nx,Gluza:2016fwh,
Peng:2012zpa,Dubovyk:2016ocz}. We describe techniques which are applied to multi-dimensional Mellin-Barnes integrals. These techniques are automatized in the Mathematica package MBnumerics, which was developed to treat numerically Feynman integrals appearing in the calculation of the electroweak two-loop corrections to the pseudo observables at the Z-boson resonance \cite{Dubovyk:2016aqv}, \cite{Dubovyk:2018rlg}.

\subsection{Linear transformation of integration variables}
In the case of the Mellin-Barnes integral the linear transformation of integration variables may lead to improvements of the  numerical integration. If we apply the variable change $z_{2}\to z_{2}-z_{1}$ to the example integral in Eq.~\eqref{eq:MBexampleseries} we get
\begin{eqnarray}
 I_{\text{0h0w14}}&=-\int\limits_{-\frac{1}{3}-i\infty}^{-\frac{1}{3}+i\infty}\frac{\dd z_{1}}{2\pi i}\int\limits_{-\frac{2}{3}-i\infty}^{-\frac{2}{3}+i\infty}\frac{\dd z_{2}}{2\pi i}\,
 \frac{(-\frac{M_{Z}^2}{s+i\delta})^{z_{1}} 
 \Gamma(-z_{1})^2 \Gamma(1 + z_{1}) \Gamma(z_{1} - z_{2}) \Gamma(-z_{2})^3  }{2 \Gamma(2 - z_{1}) \Gamma(-z_{1} - z_{2}) \Gamma(1 + z_{1} - z_{2})}\notag\\
 &\times\Gamma(1 + z_{2})\Gamma(1 - z_{1} + z_{2}).
      \label{eq:MB0h0w14finitelinearTransformation}
\end{eqnarray}
After this simple change of variable the asymptotic behavior of the Mellin-Barnes integrand has been changed. If we apply again the Stirling formula in Eq.~\eqref{eq:stirling} to the integrand in Eq.~\eqref{eq:MB0h0w14finitelinearTransformation}, and study the asymptotic behavior for $z_{1}=-\frac{1}{3}+it_{1}$, $z_{2}=-\frac{2}{3}+it_{2}$, $t_{1}\to-t$ and $t_{2}\to 0$, we find
\begin{equation}
 \CI_{\text{0h0w14}}\underset{t\to\infty}{\approx} t^{-2+2x_{2}}|_{x_{2}=-2/3},
\end{equation}
i.e. the polynomial asymptotic behavior depends only on $x_{2}$. Linear integration variable transformations give a possibility for a nontrivial cross check of the numerical evaluation of the Mellin-Barnes integrals, since the integrands have different asymptotic behavior before and after the linear transformation.

\subsection{Integrand mappings}
An obvious improvement is the application of the cotangent mapping $t=\frac{1}{\tan(-\pi d)}$, which maps the integration boundaries from $t\in[-\infty,\infty]$ to $d\in[0,1]$. We apply this mapping to a polynomial function, which gives
\begin{equation}
 \frac{1}{t^{a}}=\tan(-\pi d)^a,
 \label{eq:cotangMapping}
\end{equation}
and the Jacobian is:
\begin{equation}
\frac{\pi}{\sin(\pi d)^2},
\end{equation}
where the limits of the integrand at the boundaries of the new integration domain are:
\begin{equation}
  \lim\limits_{d\to0,d\to1}\frac{\pi\tan(-\pi d)^a}{\sin(\pi d)^2}= \begin{cases}
  \frac{1}{0}, & a<2, \\
  \pi, & a=2, \\
  0, & a>2.
  \end{cases}
\end{equation}
Compared to the cotangent mapping, a logarithmic mapping, as it is advocated in the program MB.m \cite{Czakon:2005rk}, always leads to infinities at the new  integration boundaries, which would lead to numerical instabilities.

Since we use the cotangent mapping it is mandatory to transform the integrand as follows:
\begin{equation}
 \prod_{i}\Gamma_{i}\to\exp\left(\sum_{i}\log\Gamma_{i}\right),
 \label{eq:logGammaMapping}
\end{equation}
where the key idea is that the $\log\Gamma(z_{i})$ functions grow slower than the $\Gamma(z_{i})$ functions for large values of $|z_{i}|$.

\subsection{Shifts}
If we shift the Mellin-Barnes integration variables according to
\begin{equation}
 z_{i}=x_{i}+it_{i}+n_{i},\quad n_{i}\,\in \R,
\end{equation}
the asymptotic behavior of a given Mellin-Barnes integrand may depend explicitly on the shifts $n_{i}$:
\begin{equation}
 \CI_{\text{0h0w14}}\underset{t\to\infty}{\approx} t^{-2+2x_{2}+2n_{2}}|_{x_{2}=-2/3}.
\end{equation}
It is then possible to improve the polynomial asymptotic behavior by tuning the shifts $n_{i}$.
If, by changing the values of $n_{i}$, the contour crosses some poles of the Mellin-Barnes integrand, we have to collect their residues.

\begin{figure}[tpb]
\centering
   \begin{center}
\includegraphics[width=0.48\linewidth]{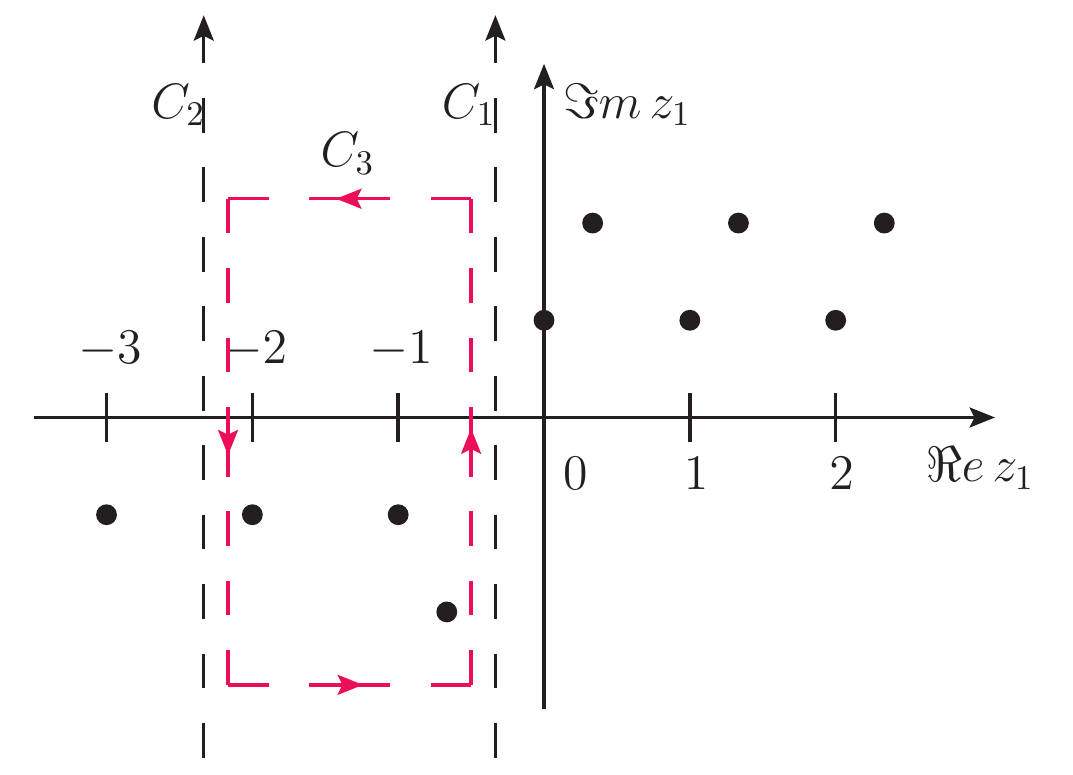}
   \end{center}
  \caption{The original contour $C_{1}$ is shifted by $n_{1}=-2$ to a contour $C_{2}$. The third contour $C_{3}$ encircles the poles to correct the shift.}
  \label{fig:contourcolor}
\end{figure}

The shifts may also be used as a method to evaluate Mellin-Barnes integrals in Minkowskian regions due to one more observation: the integral along a shifted contour may be numerically smaller by  orders of magnitude compared to the original integral.

For example, the original integral in Eq.~\eqref{eq:MB0h0w14finitelinearTransformation}, with 
$M_{Z}/\sqrt{s}=1-i\delta$, evaluated along the contour $C_1$, see Fig.~\ref{fig:contourcolor}, gives $0.3923828588857 + 0.7456388536613 i$. 
We chose $\delta = 10^{-16}$.
The shifted integral with $n_{1}=-2$, evaluated along the contour $C_2$ gives  $-0.00974965823202$. 
In addition the following equation holds:
\begin{equation}
 \int \frac{\dd\,z_{2}}{2\pi i}\int_{C_{1}}\frac{\dd\,z_{1}}{2\pi i}\CI_\text{0h0w14}  = \int\frac{\dd\,z_{2}}{2\pi i}\int_{C_{2}}\frac{\dd\,z_{1}}{2\pi i}\CI_\text{0h0w14} + \overset{\text{1 dim integrals}}{\overbrace {\int \frac{\dd\,z_{2}}{2\pi i}\left(
 \sum\limits_{z_{0}}\mathrm{Res}_{z_{0}}\CI_\text{0h0w14}\right)}}.
\label{eq:shiftequation}
\end{equation}
We have 
added to the result of the shifted integral the contributions from the residues 
of the three poles enclosed by the contour $C_{3}$. Upon integrating them over $z_{2}$, their sum is $0.402132517117807+ 0.745638853661318 i$. 
In general, shifting the contour of an $n$-fold Mellin-Barnes integral will yield residue terms, which will be $(n-1)$-fold Mellin-Barnes integrals and hence simpler to evaluate.

\section{Nontrivial example}

\begin{figure}[htpb]
\centering
   \begin{center}
\includegraphics[width=0.48\linewidth]{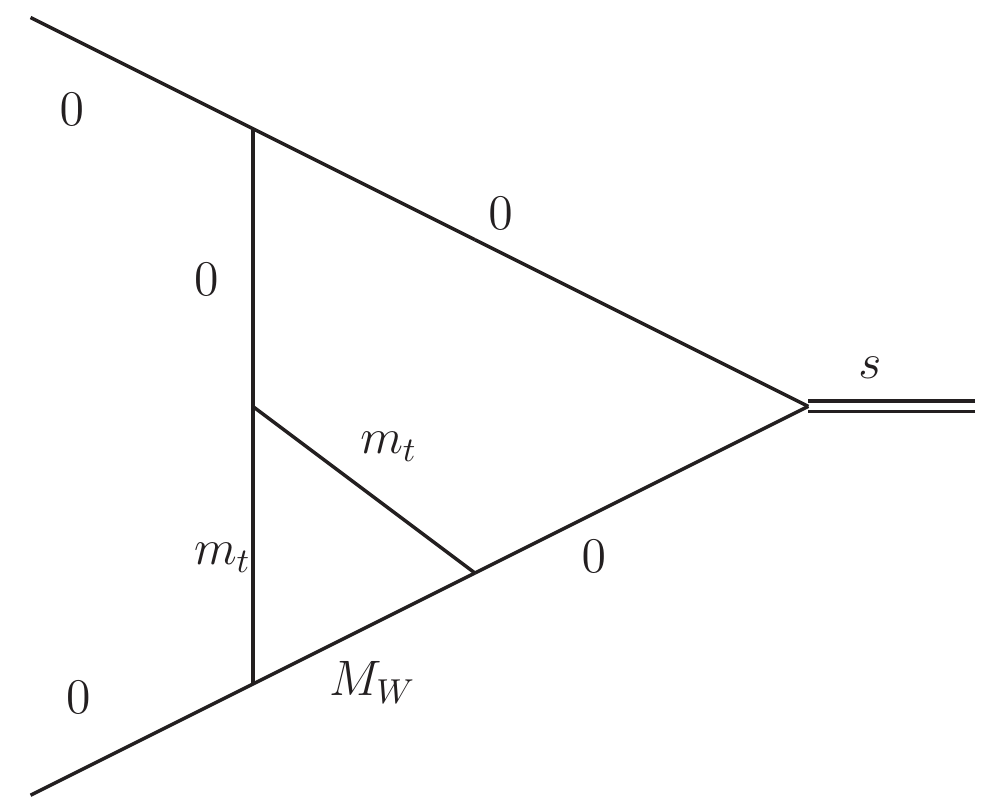}
   \end{center}
  \caption{This Feynman integral depends non-trivially on the scales $s$, $M_{W}$ and $m_{t}$.}
  \label{fig:soft}
\end{figure}

The Feynman integral shown in Fig.~\ref{fig:soft},
\begin{equation}
\int\frac{{\mathrm d}^{D} k_{1}}{i\pi^{D/2}}\frac{{\mathrm d}^{D} k_{2}}{i\pi^{D/2}}\,\frac{\exp(2\epsilon \gamma_{E})M_{Z}^{2+2\epsilon}(k_{2}p_{2})}{(k_{1})^{2}((k_{1}-k_{2})^{2} - m_{t}^2) ((k_{2})^{2} - M_{W}^2) ((k_{1}+p_{1})^{2})((k_{2}+p_{1})^{2} - m_{t}^2)(k_{1}+p_{1}+p_{2})^{2}},
\label{eq:soft}
\end{equation} 
depends in a nontrivial way on the scalar product $k_{2}p_{2}$ in the numerator.
We evaluate this integral with MBnumerics, which implements the method of the shifts. The results are collected in Tab.~\ref{tab:MBnumerics}. The scales are fixed to: 
$s=M_{Z}^2+i\delta$, $M_{Z}=91.1876\; \mathrm{GeV}$, 
$M_{W}=80.385\; \mathrm{GeV}$ and $m_{t}=173.2\; \mathrm{GeV}$.
\begin{table}[tpb]
\renewcommand{\arraystretch}{1.0}
\begin{center}
\begin{tabular}{lll}
Method & Numerics& \\\hline
    MB          & $0.0602664865576999\,\epsilon^{-2}$& \\
    SD - 90 Mio & $0.06026648655\,\epsilon^{-2}$& \\\hline
    MB          & $(-0.0315124890$ & $+0.1893327514i)\,\epsilon^{-1}$\\
    SD - 90 Mio & $(-0.031512481$  & $+0.189332716i)\,\epsilon^{-1}$\\\hline
    MB        &$(-0.22823186755$& $-0.08824794573i)+\CO(\epsilon)$\\
    SD - 90 Mio &$(-0.2282265$ & $-0.0882459 i)+\CO(\epsilon)$\\\hline
  \end{tabular}
  \end{center}
  \caption{The numbers labeled MB are evaluated with MBnumerics. The numbers labelled with
SD are evaluated with SecDec v.3 \cite{Borowka:2015mxa}.}
  \label{tab:MBnumerics}
\end{table}
In this example the Mellin-Barnes integral representation is at most a three-dimensional integral. With the sector decomposition approach the Feynman integral \eqref{eq:soft} is five-dimensional. If one can find a Mellin-Barnes integral representation whose integration dimension is smaller than or equal to that of the sector decomposition representation,
the method of shifts turns out to be very successful to compute numerically Feynman integrals in Minkowskian regions.

\section*{Acknowledgments}
We would like to thank Ayres Freitas and Janusz Gluza for fruitful discussions. We enjoyed the opportunity to complete together with them the calculation of the electroweak two-loop
corrections to the Z-boson resonance physics. 
The work of I.D. is supported by a research grant of Deutscher Akademischer Austauschdienst (DAAD) and is supported partly by the Polish National Science Centre (NCN) under the Grant Agreement 2017/25/B/ST2/01987.
The work of T.R. is supported in part by an Alexander von Humboldt Polish Honorary Research Fellowship.
This project has received funding from the European Research Council (ERC) under the European Union's Horizon 2020 research and innovation programme under grant agreement No 647356 (CutLoops). This work is supported by Graduiertenkolleg 1504 "Masse, Spektrum, Symmetrie" of Deutsche Forschungsgemeinschaft (DFG). 

%

\bibliographystyle{JHEP.bst}
\bibliography{mtools}

\end{document}